\documentclass[
    reprint,
    superscriptaddress,
    amsmath,amssymb,
    aps,
    prd,
    nofootinbib
]{revtex4-2}
 
\usepackage{subcaption}
\usepackage{graphicx}
\usepackage{float}
\usepackage{dcolumn}
\usepackage{bm}
\usepackage{empheq}
\usepackage{xcolor}
\definecolor{lightred}{RGB}{255, 220, 220}
\usepackage{soul}
\usepackage{xcolor}
\definecolor{lightred}{RGB}{255, 200, 200}
\sethlcolor{lightred} 
\usepackage[colorlinks=true,linkcolor=blue,urlcolor=blue,citecolor=blue]{hyperref}

\begin{document}

\title{Quasiparticle Modeling of Strange Quark Stars with In-Medium QCD Effects at Finite Temperature and Strong Magnetic Fields}

\author{Yogesh Kumar}
\affiliation{Department of Physics, Hansraj College, University of Delhi, Malka Ganj, Delhi-110007, India}
\author{Mihir Arora}
\affiliation{Department of Physics, Hansraj College, University of Delhi, Malka Ganj, Delhi-110007, India}
\author{Manas Lohani}
\affiliation{Department of Physics, Hansraj College, University of Delhi, Malka Ganj, Delhi-110007, India}
\author{Poonam Jain}\thanks{Corresponding Author:pjain$\_$phy@aurobindo.du.ac.in}
\affiliation{Department of Physics, Sri Aurobindo College, University of Delhi, Malviya Nagar, Delhi-110017, India}
\author{Pargin Bangotra}
\affiliation{Department of Physics, Netaji Subhas University of Technology, Dwarka, Delhi, India}
\author{Rajeev}
\affiliation{Department of Computer Science, GNIOT Institute of Professional Studies, Greater Noida, U.P., India}
\date{\today}

\begin{abstract}

This work investigates the properties of spin-polarized strange quark matter (SQM) and the resulting structure of strange quark stars (SQSs) within a thermodynamically self-consistent quasiparticle framework. By developing a dynamic bag constant intrinsically coupled to the medium-dependent reduction of quark effective masses, we ensure strict internal thermodynamic consistency. This coupling gives rise to a distinct non-monotonic behavior in the speed of sound, significantly stiffening the Equation of State (EoS) at high densities. Consequently, our model predicts maximum stable gravitational masses up to $2.19 \,M_\odot$. These results demonstrate that hot, magnetized SQM provides a robust and viable mechanism to explain ultra-massive pulsars, such as PSR J0740$-$6620 and PSR J0952$-$0607, without requiring unrealistic thermal or structural assumptions.
\end{abstract}

\maketitle

\section{Introduction}
\label{sec:intro}

The study of compact stars offers a unique window into the dynamics of matter under extreme conditions, sitting at the crossroads of nuclear low-energy physics, particle high-energy physics, and general relativity. These objects provide one of the few natural laboratories where densities far exceeding those of atomic nuclei can be explored. Among the many theoretical ideas proposed in this context, the Strange Matter Hypothesis \cite{Bodmer:1971} stands out as particularly compelling. It suggests that Strange Quark Matter (SQM) may constitute the true ground state of strongly interacting matter. If correct, this would imply the existence of Strange Quark Stars (SQSs).

Early theoretical developments in this field were largely based on the MIT bag model \cite{Chodos:1974,Chodos:1974b}, which provides a simple but effective way to explain quarks that move freely inside a localized hadronic volume bound by an external vacuum pressure. This model has played a foundational role in establishing how quark matter can support stable stellar configurations and continues to serve as an important baseline for understanding compact objects.
 
Building on this framework, subsequent studies explored more realistic physical conditions, including the effects of strong magnetic fields and spin polarization. These investigations demonstrated that magnetic fields and spin alignment can significantly influence the Equation of State (EoS) and, consequently, the macroscopic properties of compact stars \cite{Chakrabarty:1996,Broderick:2000,Bordbar:2011}. Many of these analyses were carried out under simplifying assumptions such as zero temperature, rest quark masses, and a fixed bag constant \cite{Zhang:2024_Review,Bordbar:2011}.

The inclusion of finite temperature marked an important step forward, enabling the study of hot, magnetized, and spin-polarized matter in astrophysical environments \cite{Patra:1996,Bordbar:2013}. Despite this progress, many approaches continue to treat the vacuum pressure phenomenologically, often employing density-dependent parametrizations to reproduce observational constraints. While effective, such formulations do not always maintain full thermodynamic self-consistency or adequately capture the underlying in-medium dynamics of Quantum Chromodynamics (QCD).

A more realistic description of dense matter requires incorporating medium-dependent effects. It is well known that quarks in a dense medium acquire effective masses that depend on the thermodynamic state of the system \cite{Peshier:2002}. Quasiparticle Models (QPM) provide a natural framework for implementing this behavior, treating quarks as excitations with dynamically generated masses. In the present work, we employ a thermodynamically consistent QPM in which quark masses evolve according to a running coupling determined by the relevant energy scale \cite{Peshier:2002,Chen:2022,Kumar:2018}. This scale incorporates the combined effects of baryon chemical potential, temperature, and magnetic field, allowing for a unified description of strongly interacting matter in extreme environments.

Furthermore, we address a key limitation of earlier models by constructing a variable bag constant \cite{Gorenstein:1995,Wen:2025}. Unlike conventional phenomenological prescriptions, our formulation couples the effective vacuum bag energy directly to the medium-dependent behavior of the quark quasi-particle masses, thereby ensuring thermodynamic consistency and local conservation laws throughout the phase space. An important consequence of this self-consistent coupling is the natural emergence of a non-monotonic speed of sound trajectory \cite{Peshier:2002,McLerran:2019}. By incorporating these elements under strict local charge neutrality and weak $\beta$-equilibrium, we develop a comprehensive model that bridges theoretical expectations with observations of high-mass pulsars \cite{Cromartie:2020,Romani:2022,Demorest:2010}.

In the study of compact star, it was reported that the observed magnetar's surface fields are generally consider in the range of $10^{14}-10^{15} \text{G}$   $\cite{Harley:2011,Maldonado:2020}$ while theoretical model studies claimed stronger magnetic fields in the interiors of compact star that is around $10^{18}\text{ G}$ $\cite{Bordbar:2011}$. This ultra-strong magnetic fields are believe to have a significant effect on the various properties of dense quark star. Many authors have performed the calculations of SQS with $B = 10^{18}\text{ G}$ and demonstrated that such strong fields significantly modify the thermodynamic properties of SQM $\cite{Bordbar:2011,Isayev:2014}$. In the regime of perturbative QCD, Sedaghat and co-worker $\cite{Sedaghat:2022}$ showed anisotropic effects in the presence of ultra-strong magnetic fields $B = 10^{18}\text{ G}$. Thus, in order to investigate the thermodynamic and structural features of SQM, such theoretical studies play a crucial role $\cite{Ferrer:2010}$.
\\
On the other hand, it is the fact that the realistic value of the magnetar surface field strength for the isotherms is considered as $B = 10^{15}\text{ G}$$\cite{Maldonado:2020}$. This was originally proposed by Duncan and Thompson who showed that the rotating proto-neutron star may generate field around $B = 10^{15}\text{ G}$ $\cite{Duncan:1992}$. It was further reported that such magnetic fields may affect various factors such as the thermal, magnetic, and other properties of compact stars. Above all, these two different scenario are important in the study of SQM and other dense quark star; one can distinguish between the realistic value in case of surface magnetic field considered as $B = 10^{15}\text{ G}$, and the theoretical one that can reach apround $10^{18}\text{ G}$ and more in the interior of compact stars. This distinction provides a physical basis for investigating both magnetar surface fields and ultra-strong interior fields in studies of magnetized SQM. In the view of valid observational stability and virial constraint for SQS system, we use well established range ($B = 10^{15}\text{ G}$) for the current work $\cite{Maldonado:2020}$. Also, to show the effect of varying the strength of the magnetic field on micro and macroscopic structure of a star without the danger of gravitational disruption, we have chosen the field scale range $10^{15}-10^{19} \text{G}$ for both EoS and Tolman-Oppenheimer-Volkoff (TOV) equations at temperature $T=30$ MeV.

The remainder of this paper is organized as follows: In {\ref{sec:theory}}, we present a brief model description of SQM, detailing the evaluation of effective baryonic masses {\ref{subsec:bar_mass}} alongside the fundamental thermodynamics and the formulation of the EoS {\ref{subsec:thermo}}. Section {\ref{sec:tov}} covers the stellar structure, the non-monotonic speed of sound, and numerical integration of the TOV equations. Section {\ref{comparison}} is dedicated to a comparative study and discussion, where we evaluate our primary numerical results against alternative theoretical baselines and modern observational constraints. Finally, Section {\ref{sec:conclusion}} provides the concluding remarks and outlines prospective avenues for future research.

\section{Brief Model Description of SQM}
\label{sec:theory}

The properties of SQM consisting of up ($u$), down ($d$), and strange ($s$) quarks have been investigated using various phenomenological approaches to understand the behavior of matter under extreme density \cite{Witten:1984, Farhi:1984, Schertler:1997}. We examine spin-polarized SQM where quarks are treated as fermions possessing an intrinsic quantum spin that can be oriented ``up'' ($+$) or ``down'' ($-$) relative to a quantization axis \cite{Bordbar:2011}. The degree of spin alignment is quantified by the polarization parameter $\zeta_i$ \cite{Bordbar:2013,Bordbar:2011}:
\begin{equation}
    \zeta_{i}=\frac{n_{i}^{(+)}-n_{i}^{(-)}}{n_{i}},
\end{equation}
where $n_{i} = n_{i}^{(+)} + n_{i}^{(-)}$ represents the total number density for a given quark flavor. Throughout this analysis, we assume the simplifying condition of equal polarization for all flavors, i.e., $\zeta_{u} = \zeta_{d} = \zeta_{s} \equiv \zeta$ \cite{Bordbar:2011,Bordbar:2013,Felipe:2008, Wen:2012}. Furthermore, we impose the strict conditions of local electric charge neutrality and weak $\beta$-equilibrium under the assumption that neutrinos have escaped the stellar interior. The chemical potentials of the individual quark flavors are coupled to the electron chemical potential $\mu_e$ and the baryon chemical potential $\mu_B = \mu_u + \mu_d + \mu_s$ via:
\begin{equation}
     \quad \mu_{d} = \mu_{s} = \mu_u + \mu_e = \frac{\mu_B + \mu_e}{3}.
\end{equation}
This formulation ensures that the local net electrical charge density, $n_Q = \frac{2}{3}n_u - \frac{1}{3}n_d - \frac{1}{3}n_s - n_e \equiv 0$, is perfectly preserved throughout the stellar profile, with the electron contribution explicitly incorporated into the EOS.

\subsection{Effective Baryonic Masses}
\label{subsec:bar_mass}
\begin{figure*}[t]
    \centering
    \begin{subfigure}[b]{0.48\textwidth}
        \centering
        \includegraphics[width=\textwidth]{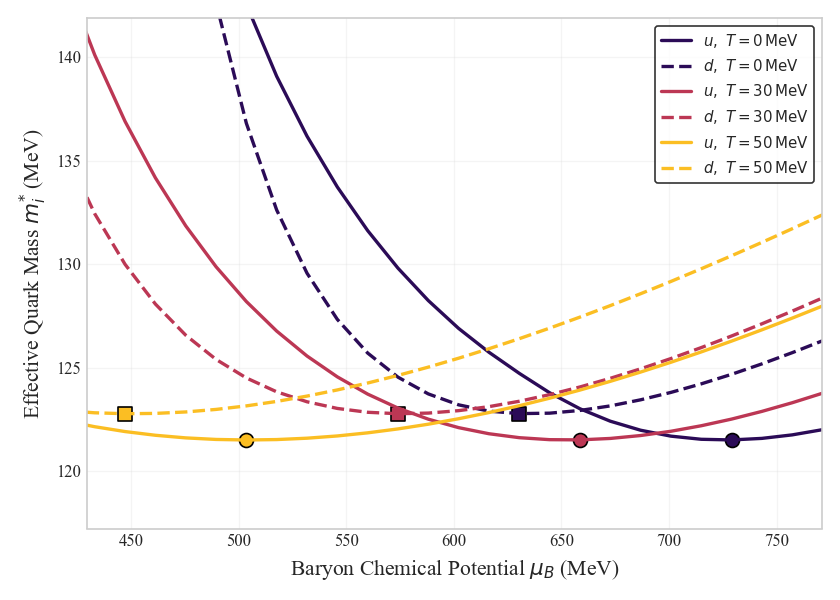}
        \caption{}
        \label{fig:eff_mass_global}
    \end{subfigure}
    \hfill
    \begin{subfigure}[b]{0.48\textwidth}
        \centering
        \includegraphics[width=\textwidth]{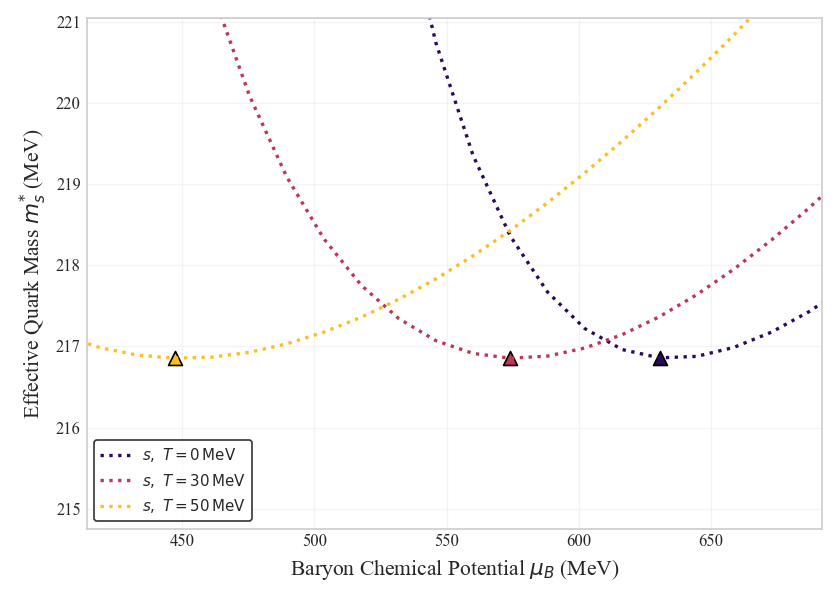}
        \caption{}
        \label{fig:eff_mass_zoom}
    \end{subfigure}
    
   \caption{Effective quark mass ($m^{*}_{i}$) as a function of baryonic chemical potential ($\mu_B$), temperature ($T$), and magnetic field ($B$) for $u$ and $d$ flavors in panel (a), and the $s$ flavor in panel (b).}
    \label{fig:1}
\end{figure*}

Building on the foundation of magnetized SQM stars \cite{Chakrabarty:1996,Broderick:2000,Bordbar:2011}, this work introduces a key modification where quark masses are not treated as constants but evolve according to the environment \cite{Peshier:2002, Schertler:1997}. The reduction of the effective mass arises naturally from the characteristic energy scale of the medium. In this framework, the renormalization scale $Q$ is taken to depend on the quark chemical potential value $\mu$ and temperature $T$, capturing the increase in interaction energy as the system moves toward the deconfined regime\cite{Peshier:2002,Kurkela:2010}:
\begin{equation}
    Q_{i} = \sqrt{\mu_{i}^2 + (\pi T)^2}.
\end{equation}
The interaction strength is governed by the running coupling constant $g^2$ \cite{Farias:2014,Wen:2025,Schertler:1997}:
\begin{equation}
    g_{i}^{2}(Q,B,T)=\frac{48\pi^{2}}{(33-2N_{f})\ln\left(\frac{Q_{i}^{2}}{Q_{min}^{2}}\right)\mathcal{C}_{mag}(B)},
\end{equation}
where $N_f = 3$ is the number of active quark flavors. Here, $Q_{min}$ represents the lower bound of the renormalization scale, ensuring the logarithmic argument remains physical. This is defined as $Q_{min} = \Lambda_{1}/\sqrt{0.8}$ \quad $\text{with} \quad \Lambda_{1} = 120 \text{ MeV}$ \cite{Wen:2025} enforcing the condition that $\mu$ must be greater than the QCD scale. The magnetic correction factor, which accounts for the modification of the strong coupling constant in the presence of an external magnetic field, is given by \cite{Farias:2014, Wen:2025}:
\begin{equation}
\mathcal{C}_{\mathrm{mag}}(B) = 1 + \alpha \ln \left( 1 + \beta\frac{ eB}{\Lambda_{2}^{2}} \right),
\end{equation}
where the scale parameters are $\alpha = 2$, $\beta = 3.27\times10^{-4}$, and $\Lambda_2 = 200~\mathrm{MeV}$ \cite{Wen:2025}. To preserve the underlying microphysical configuration, the baseline magnetic profile is restricted to a physical limit of $B = 10^{15}\text{ G}$. This specific choice complies with the empirical upper boundary for stable compact configurations, where a severe magnetic pressure dominance ($\propto B^2 / 8\pi$) would otherwise distort spherical geometry or exceed the energetic constraints defined by standard magnetar baselines $\cite{Maldonado:2020,Duncan:1992}$. 
The effective mass $m_{i}^{*}$ for each flavor is determined by the local thermodynamic scales through the following relation \cite{Schertler:1997, Wen:2025, Wen:2012}:
\begin{equation}
    m_{i}^{*}(T, \mu_i) = \frac{m_{i}}{2} + \sqrt{\frac{m_{i}^{2}}{4} + \frac{g_{i}^{2}Q_{i}^{2}}{6\pi^{2}}}
    \label{eq:effective_mass_unified}
\end{equation}
where $m_{i}$ represents the rest masses ($m_{e}=0.511$, $m_{u}=2.2$, $m_{d}=4.7$, and $m_{s}=150$ MeV), and the second term under the radical accounts for the medium-dependent mass gap arising from non-perturbative gluon-exchange interactions \cite{Wen:2025,Schertler:1997}.
The total baryonic number density of the system is given by the sum of these densities:
\begin{equation}
    n_{B} = \frac{1}{3} \sum_{i=u,d,s} n_{i}.
\end{equation}
The variation of the effective quark masses $m_i^{*}$ with parameters  $\mu_B$, $T$, and $B$ is shown in Fig.~\ref{fig:1}. At low $\mu_B$, the rapid non-linear decrease in $m_i^{*}$ signals the transition from a strongly interacting regime to deconfined quark matter. Finite temperature shifts this transition to lower $\mu_B$, indicating thermal facilitation of deconfinement. The strange quark maintains a higher mass scale ($\approx 215\text{--}230$ MeV) relative to light quarks due to its larger rest  mass. The minima in $m_i^{*}$ mark the onset of stabilized quasiparticle behavior. This evolution plays a key role in shaping the EoS.

\subsection{Thermodynamics and Equation of State}
\label{subsec:thermo}

To describe the statistical properties of spin-polarized SQM at a finite value of temperature and in the presence of a high value of magnetic field, we employ the Fermi--Dirac distribution function for each quark flavor and polarisation $p$ ($p=\pm1$ corresponding to spin-up and spin-down, respectively) \cite{Bordbar:2013,Peshier:2000,Wen:2012,Pal:2023}: 
\begin{equation}
f^{(p)}_{i}(k,T)=
\left[
\exp\left(
\frac{E_{i}(k)-\mu^{(p)}_{i}}{k_B T}
\right)
+1
\right]^{-1}
\end{equation}

Here,
\begin{equation}
E_{i}(k)=\sqrt{k^2 + {m_i^*}^2}
\end{equation}
represents the quasiparticle energy spectrum\cite{ Schertler:1997,Bordbar:2011}, where $k_{B}$ is the Boltzmann constant. The spin-dependent chemical potential is given by
\begin{equation}
\mu^{\pm}_{i}=\mu_{i} \pm\,{\mu_{0}}_{i} B ,
\end{equation}
where ${\mu_{0}}_{i}$ is the magnetic moment associated with the flavour $(\mu_{0u} = 1.852\,\mu_N$, $\mu_{0d} = -0.972\,\mu_N$, $\mu_{0s} = -0.581\,\mu_N$, $\mu_{0e} = -1.0011\,\mu_N$ and $\mu_N = 3.15245 \times 10^{-14}\text{ MeV/T})${\cite{Chakrabarty:1996}\cite{Bordbar:2013}}. The number density for each spin-polarized flavor is obtained by integrating the distribution function over momentum space\cite{Bordbar:2013,Peshier:2000,Pal:2023}: 
\begin{equation}
    n^{(p)}_{i} = \frac{g}{2\pi^2} \int_{0}^{\infty} k^2 f^{(p)}_{i}(k,T) dk,
\end{equation}
where $g=3$ represents the color degeneracy. The kinetic energy density  is derived by accounting for the contribution of all flavors, electron and spin states\cite{Bordbar:2013,Peshier:2000,Schertler:1997}:
\begin{equation}
    \epsilon_{kin} = \sum_{i=u,d,s,e} \frac{g}{2\pi^2} \int_{0}^{\infty} k^2 E_{i}(k) \left(f^{(+)}_{i}+f^{(-)}_{i}\right) dk.
\end{equation}
In a strongly magnetized environment, the interaction between the quark magnetic moments and the external magnetic field contributes to the total energy of the system. Within the spin-polarization framework, the magnetic energy density is given by \cite{Wen:2012,Bordbar:2011,Bordbar:2013}

\begin{equation}
\epsilon_{\mathrm{mag}} = - \sum_{i=u,d,s,e} n_i \mu_i \zeta_i B .
\end{equation}

The execution of our EoS relies on the explicit tracking of a  medium-dependent bag potential $B^{*}(\mu,T)$. Because the underlying quasiparticle masses are implicit functions of the surrounding density and temperature coordinates, the background vacuum baseline must accommodate these shifts to prevent internal thermodynamic contradictions.

Within this framework, the spatial evolution of $B^*(\mu, T)$ acts as a differential constraint that links the phase-space integrals directly to the mechanical pressure tensor. This variable profile ensures that the bulk particle number densities and entropy distributions remain exact partial derivatives of a single, unified characteristic potential. By canceling the anomalous mass-gradient artifacts of fixed-mass approximations, this coupled approach stiffens the EoS to support ultra-massive stellar configurations near the stability limit \cite{Pal:2023,Zhang:2026_SQM}.

\begin{empheq}{equation}
\begin{aligned}
B^{*}(\mu_u,\mu_d,\mu_s,T)
&= B_{0}
-\sum_{i=u,d,s}
\int_{(0,0)}^{(\mu_i,T)}
\frac{\partial F_{qp}}{\partial m_i^{*}}
\\
&\qquad\times
\left(
\frac{\partial m_i^{*}}{\partial \mu_i}\, d\mu_i
+
\frac{\partial m_i^{*}}{\partial T}\, dT
\right).
\end{aligned}
\end{empheq}
where $B_0= 55 \text{ MeV fm}^{-3}$ is the vacuum reference point  at $\mu_i=T=0$ and $F_{qp}$ represents the on-interacting quasiparticle free energy density.  The partial derivative with respect to the effective mass is evaluated analytically from the corresponding thermodynamic integration relation:
\begin{empheq}{equation}
\frac{\partial F_{qp}}{\partial m_i^*} = \frac{g}{2\pi^2} \int_{0}^{\infty} \frac{m_i^*}{E_{i}(k)} \left(f^{(+)}_{i}+f^{(-)}_{i}\right)  k^2 dk,
\end{empheq}
\begin{figure}[H]
    \centering
    \includegraphics[width=0.48\textwidth] {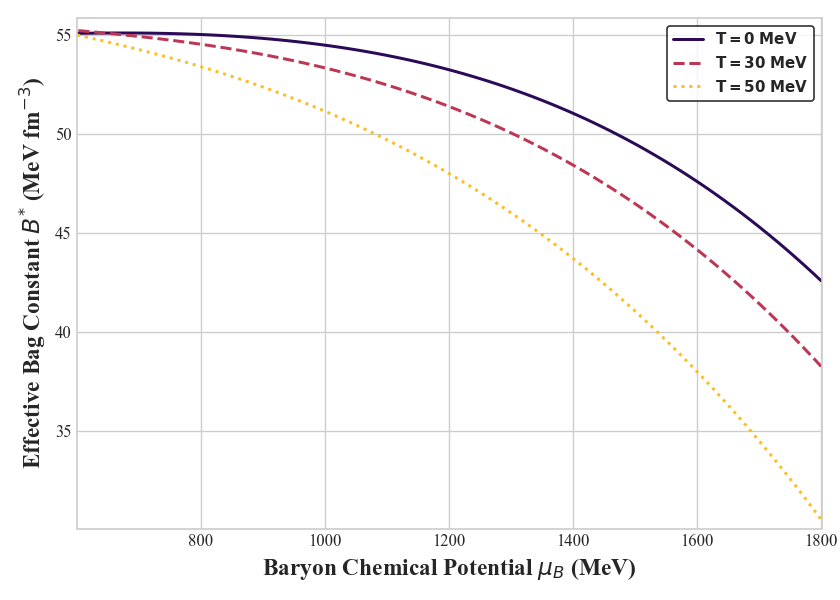}
    \caption{Evolution of the effective bag constant ($B^{*}$) as a function of baryonic chemical potential ($\mu_B$) for varying temperatures ($T$) and magnetic field ($B$).}
    \label{fig:eff_bag}
\end{figure}

As illustrated in Fig.~\ref{fig:eff_bag}, $B^{*}$ decreases as the system approaches a more relativistic regime. Unlike phenomenological Gaussian parametrizations, this formulation directly couples the vacuum pressure to the running coupling, providing a physically motivated description of in-medium effects. The rapid reduction in $B^{*}$ effectively stiffens the EoS, supplying the necessary repulsive contribution to support pulsars with masses exceeding $2\,M_{\odot}$ while remaining consistent with causality.

\begin{figure}[htbp]
    \centering
    \includegraphics[width=0.5\textwidth]{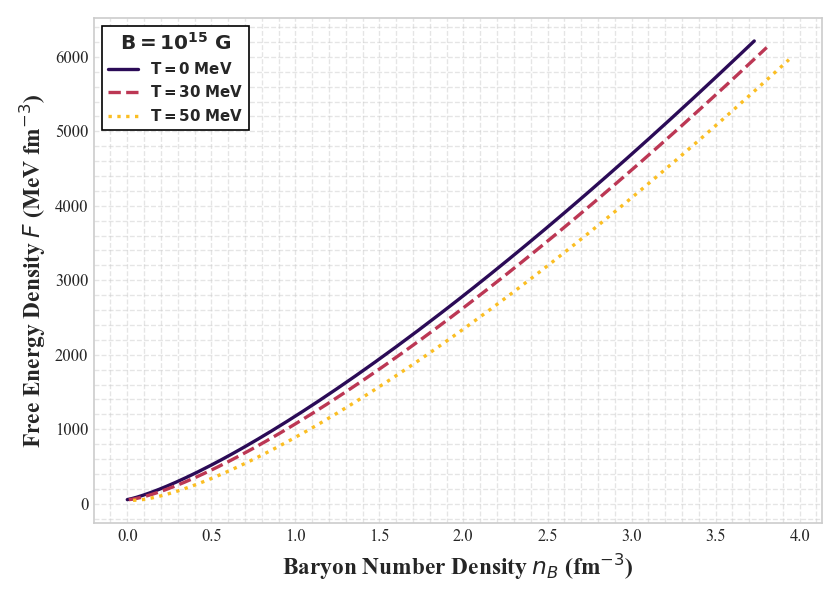}
    \caption{Helmholtz free energy density ($F$) of  SQM as a function of baryon number density ($n_B$) for $B=10^{15}$ G.}
    \label{fig:free_energy_F}
\end{figure}

\begin{figure}[H]
    \centering
    \includegraphics[width=0.5\textwidth]{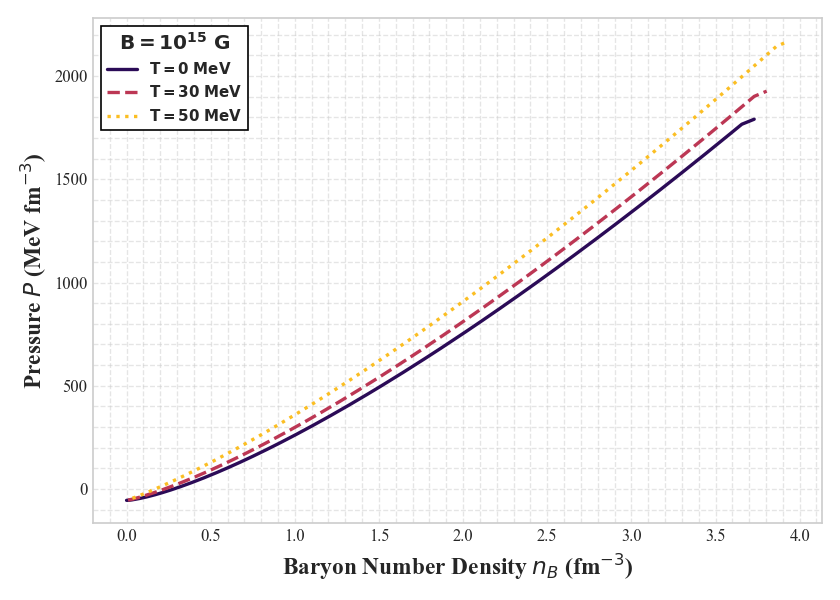}
    \caption{Pressure ($P$) of magnetized SQM as a function of baryon number density ($n_B$) for varying $T$.}
    \label{fig:pressure_P}
\end{figure}

The total energy density ($\epsilon_{tot}$) is the summation of the kinetic, magnetic, vacuum components and Maxwell field term:
\begin{equation}
    \epsilon_{tot} = \epsilon_{kin} + \epsilon_{mag} + B^{*} + \frac{B^{2}}{8\pi}
\end{equation}
To determine the pressure at finite temperature, we calculate Helmholtz free energy ($F_{tot}$)\cite{Bordbar:2013,Bordbar:2011}:
\begin{equation}
    F_{tot} = \epsilon_{tot} - TS_{tot},
\end{equation}
where the net entropy density $S_{tot}$ is derived from the Fermi-Dirac statistics\cite{Bordbar:2013,Bordbar:2011}:
\begin{equation}
    S_{tot} = \sum_{i,p} \frac{-g}{2\pi^2} \int_{0}^{\infty} k^2 [f^{(p)}_{i} \ln f^{(p)}_{i} + (1-f^{(p)}_{i}) \ln (1-f^{(p)}_{i})] dk.
\end{equation}
Finally, the thermodynamic pressure ($P$) is defined through the relation\cite{Bordbar:2013,Wen:2012}:
\begin{empheq}{equation}
    P = \sum_{i=u,d,s,e}\left(n _i \frac{\partial F_{i}}{\partial n_{i}} -F_{i}\right)
\end{empheq}
Where $F_{i}$ is the flavour-specific free energy. 
The thermodynamic evolution of the magnetized SQM configuration under a baseline magnetic field is illustrated across distinct isotherms in Fig.~\ref{fig:pressure_P} and Fig.~\ref{fig:free_energy_F}. At low baryon number densities ($n_B \lesssim n_0$), the pressure and Helmholtz free energy trajectories cluster closely, demonstrating that the microphysics in this regime are predominantly anchored by the quantum degeneracy pressure of the quark phase space and the baseline vacuum energy. Beyond the nuclear saturation scale ($n_B > n_0$), the system profiles for different temperatures fan out and exhibit quite distinct behaviors. This geometric separation reflects a systematic, temperature-driven inflation of the macroscopic pressure at a fixed baryon density, signaling an effective stiffening of the underlying EoS.

As temperature increases, the enhanced kinetic phase space and the self-consistent adjustment of the bag constant amplify the outward mechanical pressure. This behavior is further corroborated by the monotonic trajectories of  $F$ shown in Fig.~\ref{fig:free_energy_F}. Notably, at a fixed baryon density, the highly thermalized isotherm ($T = 50\text{ MeV}$) maintains a lower free energy density baseline relative to the cold matter counterpart ($T = 0\text{ MeV}$). This behavior indicates that thermal contributions effectively minimize the energetic barrier required to maintain deconfined SQM configurations, optimizing the thermodynamic stability of the plasma medium.

\begin{figure}[htbp]
    \centering
    \includegraphics[width=0.5\textwidth]{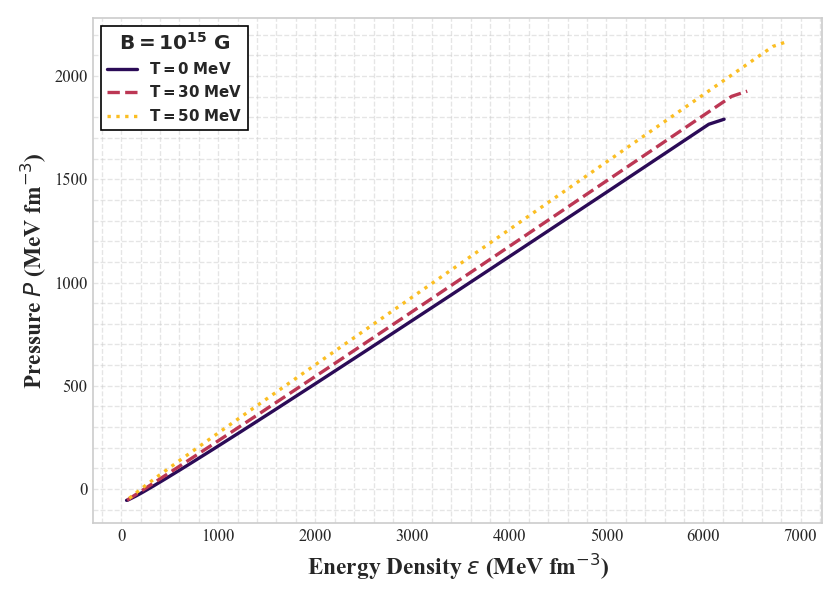}
    \caption{EoS represented as pressure $P$ versus energy density $\epsilon$ for varying $T$.}

    \label{fig:3}
\end{figure}

To evaluate how these microscopic variations dictate macroscopic stellar equilibria, Fig.{~\ref{fig:3}} maps the resulting EoS directly onto the $P$--$\epsilon$ plane. The high-density regimes of the $P(\epsilon)$ trajectories display a highly linear behavior characteristic of strongly interacting, deconfined quark plasma. Crucially, enforcing strict local charge neutrality and weak $\beta$-equilibrium through the leptonic channel yields an upward vertical shift in the finite-temperature curves relative to the cold matter baseline. Consequently, for a given $\epsilon$, highly thermalized states generate a significantly higher mechanical pressure tensor. 

\begin{figure}[htbp]
    \centering
    \includegraphics[width=0.5\textwidth]{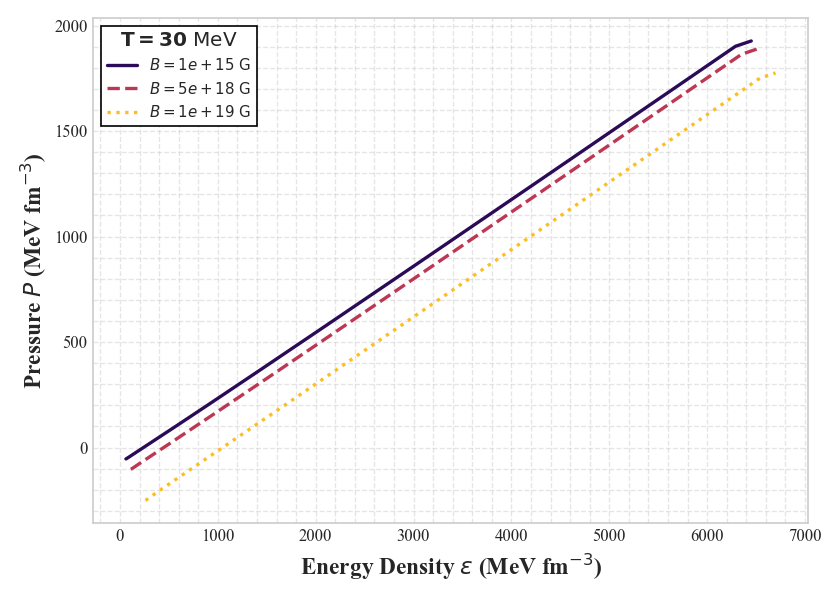}
    \caption{EoS represented as pressure $P$ versus energy density $\epsilon$ for $T=30$ MeV.}
    \label{fig:eos_B}
\end{figure}

Fig.{~\ref{fig:eos_B}} isolates the EoS exclusively along the intermediate $T = 30\text{ MeV}$ isotherm under varying field intensities. In the high energy density regime ($\epsilon > 1400\text{ MeV~fm}^{-3}$), the $P(\epsilon)$ trajectories demonstrate a highly organized, upward vertical migration with increasing magnetic field strength. This field-driven elevation in pressure at a fixed energy density signifies an additional magnetic stiffening mechanism, primarily dominated by the longitudinal magnetic field contribution. Crucially, the absence of oscillating crossover behaviours across these distinct field profiles highlights the structural stability of the dynamic vacuum integration, confirming that the dynamic tracking of $B^*$ accommodates large Lorentz forces without breaking thermodynamic continuity.

The enhanced pressure support generated by the high-temperature quasiparticle medium directly stiffens the EoS, providing the necessary mechanical resistance to counteract extreme gravitational fields. These trends show that the  stability of ultra-massive compact objects is closely tied to the high-temperature response and structural self-consistency of dense QCD matter.

\section{Stellar Structure and TOV Equations}
\label{sec:tov}

To investigate the macroscopic properties of SQSs derived from our quasiparticle EoS, we model the stars as spherically symmetric distributions of static matter in the framework of general relativity. The hydrostatic equilibrium of the star is governed by the TOV equations\cite{Tolman:1939,Oppenheimer:1939}:

\begin{equation}
\begin{split}
\frac{dP}{dr} = & -\frac{G m(r) \epsilon(r)}{r^{2}} \left( 1 + \frac{P(r)}{\epsilon(r)} \right) \\
& \times \left( 1 + \frac{4\pi r^{3} P(r)}{m(r)} \right) \left( 1 - \frac{2G m(r)}{r} \right)^{-1}
\end{split}
\label{eq:tov_split}
\end{equation}

where $P(r)$ and $\epsilon(r)$ represent the pressure and energy density at a radial distance $r$, respectively. The gravitational mass $M(r)$ contained within a sphere of radius $r$ is determined by the mass continuity equation:

\begin{equation}
    \frac{dM}{dr} = 4\pi r^{2} \epsilon(r).
\end{equation}

By integrating these equations numerically from the stellar center ($r=0$) with a given central pressure $P_c$ out to the surface ($r=R$), where the pressure vanishes ($P(R)=0$), we obtain the total gravitational mass $M$ and total radius $R$ of the star. 
\newline
A crucial requirement for any physical EoS is that it must satisfy the conditions of causality and mechanical stability. The speed of sound in the stellar medium, $v_s$, is defined by the derivative of pressure with respect to energy density\cite{Glendenning:2000}:

\begin{equation}
    \left( \frac{v_s}{c} \right)^2 = \frac{dP}{d\epsilon}.
\end{equation}

In the high-density limit of deconfined quark matter, the speed of sound is expected to approach the conformal limit of $v_s^2/c^2 = 1/3$ \cite{Kurkela:2010}. Deviations from this limit are therefore of considerable interest, as they provide insight into the underlying interaction dynamics.

\begin{figure}[H]
    \centering  
    \includegraphics[width=0.5\textwidth]{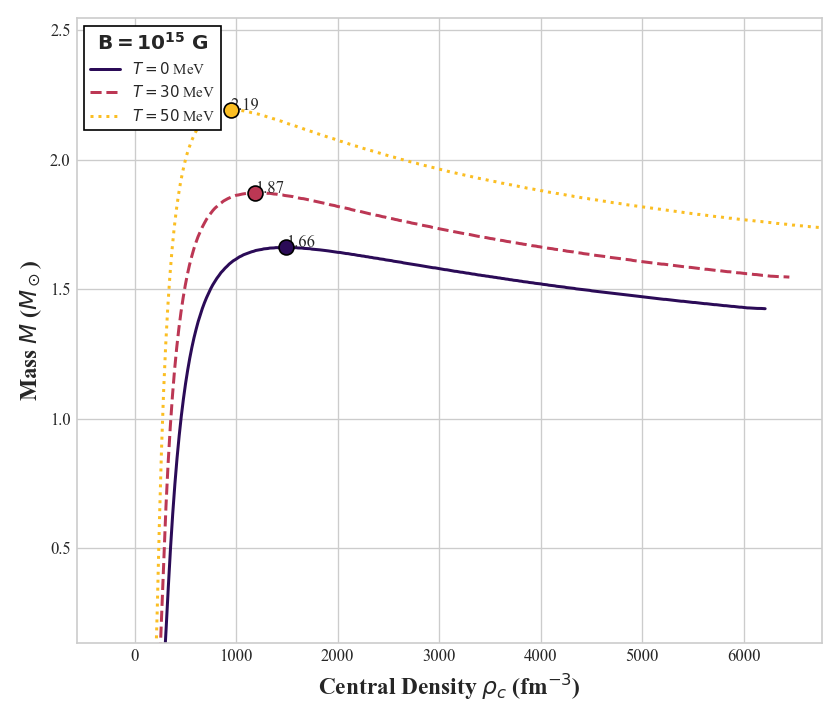}
    \caption{Stellar mass ($M$) as a function of central energy density ($\epsilon_c$) for SQSs for $B=10^{15}$ G.}

    \label{fig:mass_central_density}
\end{figure}
\begin{figure}[H]
    \centering
    \includegraphics[width=0.5\textwidth]{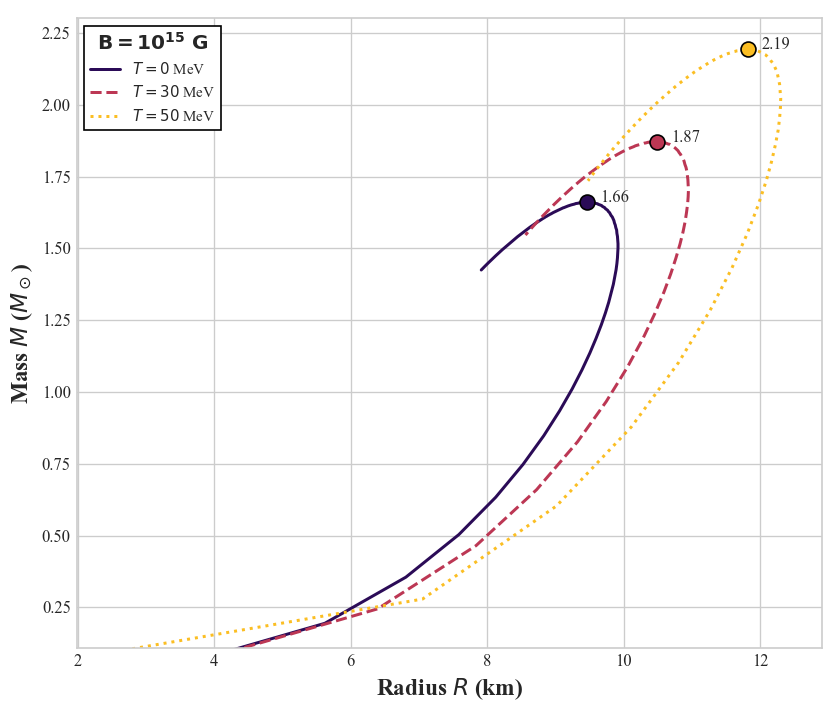 }
    \caption{Stellar mass ($M$) as a function of radius ($R$) of SQS for $B=10^{15}$G}

    \label{fig:mass_radius}
\end{figure}
\begin{figure}[H]
    \centering
    \includegraphics[width=0.5\textwidth]{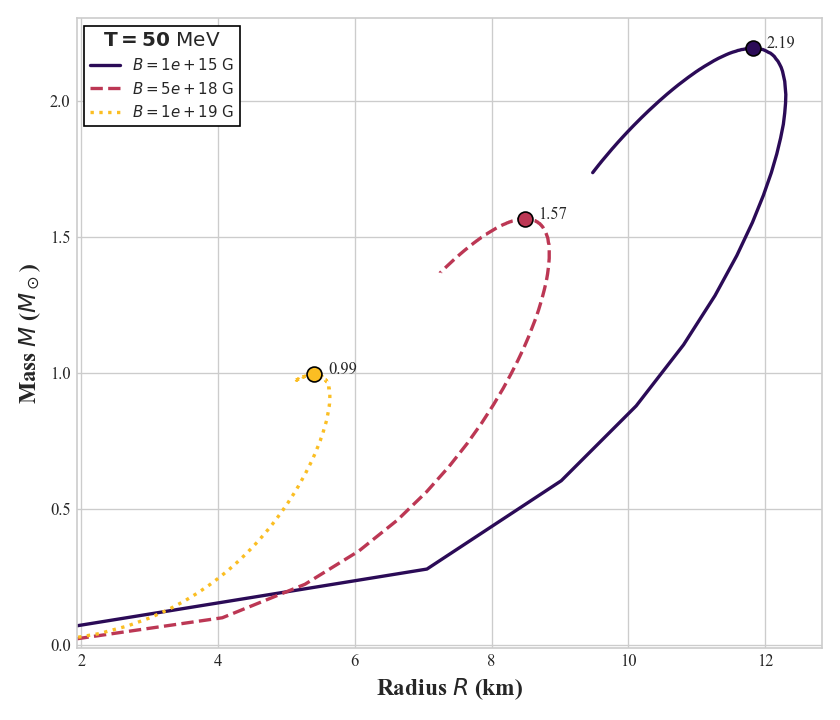}
    \caption{Stellar mass ($M$) as a function of radius ($R$) of SQS for $T=50$ MeV.}

    \label{fig:mass_radius_mag}
\end{figure}
\begin{table}[H]
    \centering
    \caption{Maximum mass configurations obtained from the present QPM for varying $T$ and $B = 10^{15}\text{ G}$ .}
    \label{tab:TOV_QPM}
    \renewcommand{\arraystretch}{1.2}
    \begin{tabular}{ccc}
        \hline
        $T$ (MeV) & $M_{\max}$ ($M_\odot$) & $R_{\max}$ (km) \\ \hline
        0  & 1.66& 9.48\\
        30 & 1.87& 10.46\\
        50 & 2.19& 11.82\\ \hline
    \end{tabular}
\end{table}
\begin{table}[H]
    \centering
    \caption{Maximum mass configurations obtained from the present QPM for varying $B$ and  $T = 50\text{ MeV}$ .}
    \label{tab:TOV_QPM_2}
    \renewcommand{\arraystretch}{1.2}
    \begin{tabular}{ccc}
        \hline
        $B$ (G)& $M_{\max}$ ($M_\odot$) & $R_{\max}$ (km) \\ \hline
        $10^{15}$& 2.19& 11.82\\
        $5\cdot 10^{18}$& 1.57& 8.50\\
        $10^{19}$& 0.99 & 5.41\\ \hline
    \end{tabular}
\end{table}
\begin{table}[H]
    \centering
    \caption{Maximum mass configurations obtained from the holographic Einstein--Maxwell--Dilaton (EMD) framework \cite{Zollner:2023}.}
    \label{tab:TOV_EMD}
    \renewcommand{\arraystretch}{1.2}
    \begin{tabular}{ccc}
        \hline
        $T$ (MeV) & $M_{\max}$ ($M_\odot$) & $R_{\max}$ (km) \\ \hline
        50  & 0.41 & 6.04 \\
        75  & 0.97 & 7.90 \\
        150 & 2.03 & 12.96 \\
        200 & 1.06 & 6.01 \\ \hline
    \end{tabular}
\end{table}
\begin{figure}[H]
    \centering
    \includegraphics[width=0.5\textwidth]{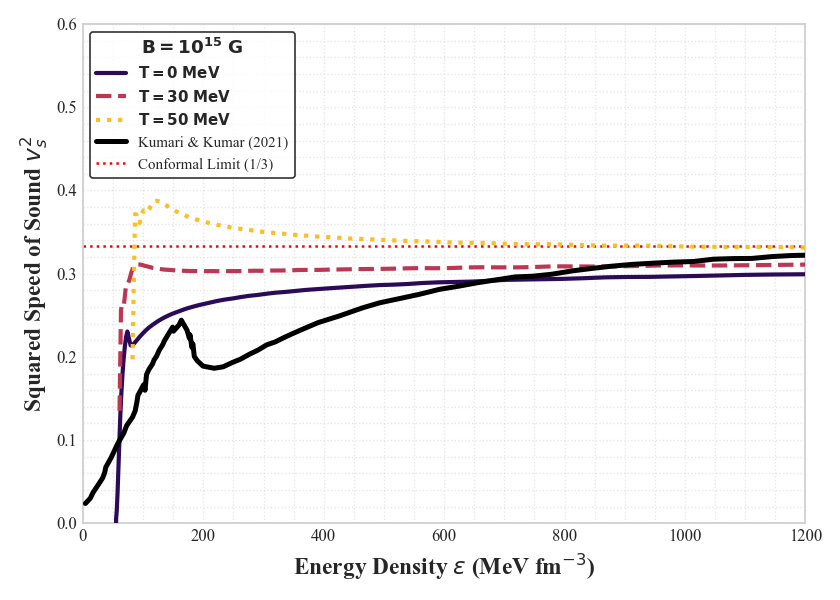}
    \caption{The squared speed of sound $v_s^2$ as a function of energy density $\epsilon$ for a fixed magnetic field $B = 10^{15}$~G. The bold solid black line provides a benchmark using the non-magnetized strange matter results from Kumari and Kumar \cite{Kumari:2021} at $T=10$ MeV. The dotted red line denotes the conformal limit ($v_s^2 = 1/3$).}
    \label{fig:speed_of_sound_final}
\end{figure}

\noindent To connect these microscopic velocity signatures with macroscopic stellar geometries, the structural profiles of the SQSs obtained via the numerical integration of the TOV equations are displayed in Figs.{~\ref{fig:mass_central_density}}, {\ref{fig:mass_radius}}, and {\ref{fig:mass_radius_mag}}. The mass--central density ($M$--$\epsilon_c$) relations reveal a migration of the global stellar stability limit (defined by the variational terminal coordinate $dM/d\epsilon_c = 0$) under changing thermodynamic configurations. For the cold matter baseline ($T=0$~MeV) at a background field of $B = 10^{15}$~G, the model yields a maximum gravitational mass of $1.66\,M_{\odot}$ with a corresponding compact radius of $9.48$~km, as summarized in Table{~\ref{tab:TOV_QPM}}. These baseline parameters demonstrate strong structural convergence with the magnetized configurations reported by Sedaghat \textit{et al.} {\cite{Sedaghat:2022}} within perturbative QCD limits, verifying the stability of our underlying numerical infrastructure before the activation of thermal channels. When finite-temperature corrections are introduced, thermal pressure inflation systematically stiffens the core, elevating the maximum supported stellar mass to $1.87\,M_{\odot}$ at $T=30$~MeV, and further to $2.19\,M_{\odot}$ at $T=50$~MeV. Crucially, this mass enhancement is accompanied by a monotonic shift of the critical stability point toward lower central energy densities, indicating that thermal phase-space expansions actively optimize mechanical resistance against gravitational collapse at lower structural densities.

\noindent The physical origin of this behavior is further elucidated by the evolution of the squared speed of sound $v_s^2$ presented in Fig.~\ref{fig:speed_of_sound_final}. We provide a direct comparison with the non-magnetized strange matter results of Kumari and Kumar \cite{Kumari:2021} at $T=10$~MeV. Under a moderate-field of $B = 10^{15}$~G alongside strict local charge neutrality yields a much smoother, continuous trajectory. While the high-temperature isotherm ($T=50$~MeV) exhibits a transient non-monotonic peak near the low-density phase-activation threshold, the profiles asymptotically flatten at higher energy densities. Rather than aggressively violating the conformal baseline, the speed of sound trajectories across all evaluated temperatures remain stable and asymptote toward the conformal limit ($v_s^2 \rightarrow 1/3$) from below at high densities.

\noindent To isolate the exact macro-structural impact of pure magnetic field variations at high temperatures, Fig.{~\ref{fig:mass_radius_mag}} maps the $M$--$R$ trajectories exclusively along the highly thermalized $T=50$~MeV isotherm for varying field scales, with quantitative profiles compiled in Table{~\ref{tab:TOV_QPM_2}}. Because SQSs are self-bound configurations governed by vacuum boundaries rather than gravitational wrapping alone, the trajectories initially follow the characteristic scaling law $R \propto M^{1/3}$ in the low-mass regime. As the background field scales up to extreme magnetar bounds ($B = 10^{19}$~G), the maximum stellar mass shifts from  $2.19\,M_{\odot}$ to $0.99\,M_{\odot}$, while the corresponding maximum radius shrinks from $11.82$~km to $5.41$~km. This inward migration of the terminal mass coordinate is directly driven by the maxwell field term ($\propto B^2 / 8\pi$). The smooth, non-intersecting nested topology of these curves confirms that the underlying integration of the dynamic vacuum potential $B^*$ maintains flawless mathematical continuity under strong Lorentz forces.

\begin{figure*}[t]
    \centering
    \includegraphics[width=1.0\textwidth]{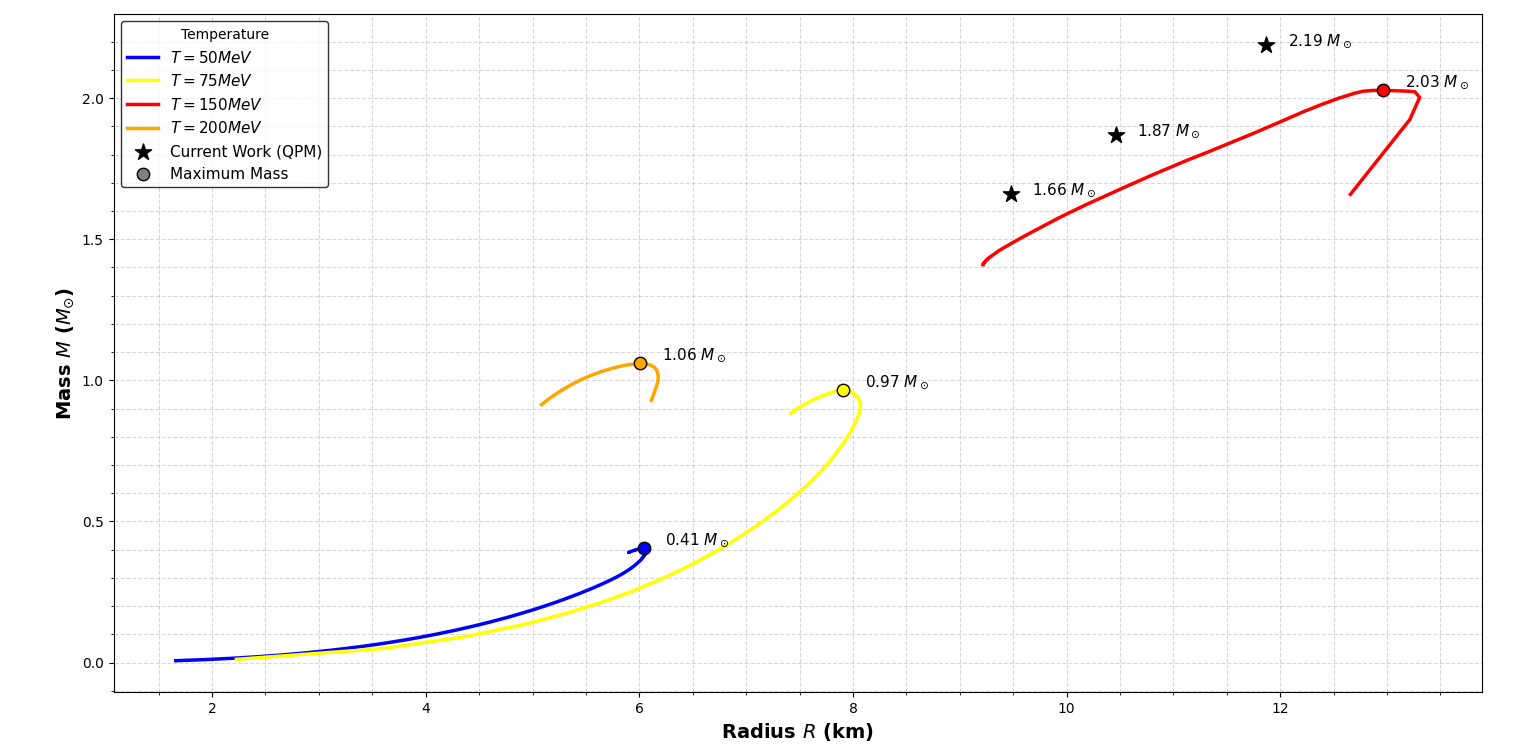}
    \caption{Mass–radius ($M$–$R$) relations for SQSs. 
The solid colored curves are obtained from EMD EoS data at different temperatures \cite{Zollner:2023}. 
The star symbols ($\star$) denote the maximum-mass configurations predicted by the present magnetized (current work) QPM, as listed in Table~\ref{tab:TOV_QPM}.}
    \label{fig:mass_radius_comparison}.
    \label{fig:holographic_mr}
\end{figure*}

\section{Comparative Study and Discussion}
\label{comparison}

\noindent In this section, we contrast our present magnetized QPM model with the holographic EoS using the EMD theory \cite{Zollner:2023}. As shown in Fig.~\ref{fig:holographic_mr}, our EoS prediction gives a distinctive benchmark to SQS configurations. The maximum mass predicted by the holographic EoS is approximately $2.03\,M_{\odot}$ at ultra-high temperature ($T=150$ MeV), whereas the corresponding value of our magnetized QPM model is $2.19\,M_{\odot}$, which is obtained at a relatively lower temperature regime of $T=50$ MeV and weak magnetic field.

\noindent The necessity for such extreme temperatures in the EMD framework arises from its reliance on the dilaton-field coupling and the Hawking temperature of the 5D black hole to mimic QCD thermodynamics. As summarized in Table~\ref{tab:TOV_EMD}, the maximum mass initially increases with temperature up to $T=150$~MeV, however, as the temperature increases further in the EMD framework, the stability limit undergoes a sharp reduction to approximately $1.06\,M_{\odot}$ at $T=200$ MeV \cite{Zollner:2023}. This signifies a thermal softening of the holographic EoS as it approaches the deconfined phase, marking a physical ``breaking point'' for the framework's support of massive stars.

\noindent In contrast, the contrasting feature between the stability curves in Fig.~\ref{fig:holographic_mr} is quite evident, with the maximum mass solutions in the present model marked with $\star$, being well above the holographic $M-R$ curves. It is clear that the effect of strong magnetic fields along with medium-modified quasiparticle masses leads to a stiff EoS compared to the dilaton-induced back reaction of the EMD model, effectively bypassing the need for extreme thermal excitation and preventing the premature instability observed in purely thermal or holographic models.

\noindent The key feature of the quantitative summary shows how effective the magnetic QPM is in fulfilling observation restrictions. While the holographic model would need super high temperatures ($T > 100$ MeV) to reach the $2\,M_{\odot}$ threshold, our model satisfies PSR J0952$-$0607 ($2.35 \pm 0.17\, M_{\odot}$) at a more plausible $T=50$ MeV. It can be concluded from here that the presence of magnetic fields greatly contributes to thermal fields. This synergy ensures that SQS remain stable at higher masses and lower temperatures than expected in models relying solely on holographic or purely thermal stiffening mechanisms.

\section{Conclusion}
\label{sec:conclusion}

\noindent In this work, we have presented a comprehensive investigation into the thermodynamic properties of magnetized SQM and its structural implications for the stability of ultra-massive SQSs. By integrating medium-dependent quark effective masses with an evolving bag constant, our framework captures the intricate coupling between quasiparticle degrees of freedom and the non-perturbative QCD vacuum.

\noindent The primary novelty of this study lies in the demonstration that the synergy between a magnetar-strength magnetic field and finite-temperature effects acts as a decisive driver for stellar stability, effectively stiffening the EoS beyond the limits of standard hadronic or non-magnetized quark models. Our results indicate a substantial increase in the maximum gravitational mass, rising from $1.66\,M_{\odot}$ in the cold regime to $2.19\,M_{\odot}$ at $T = 50\text{ MeV}$ under a moderate-field configuration. This finding is physically rooted in the rapid chiral mass reduction of quasiparticles, which when coupled with magnetic pressure provides a robust repulsive response capable of sustaining extreme gravitational loads.

\noindent A key indicator of consistency within our theory is the presence of a non-monotonic squared sound speed $(v_s^2)$. This property offers sufficient stiffness to support some of the largest known pulsars like PSR J0740+6620 and the recently discovered PSR J0952$-$0607. In addition, our comparison of the QPM under magnetization conditions with holographic EMD models reinforces the importance of this paper. We show that magnetized QPM fulfills these demanding criteria at realistic temperatures, while non-magnetized holographic models need extremely high temperatures $(T > 100$ MeV).

\noindent The relevance of this paper extends to the interpretation of high-mass compact objects in the era of multi-messenger astronomy. The ability of thermalized, magnetized SQSs to reach masses exceeding $2\,M_{\odot}$ suggests that secondary components in high-mass merger events, such as the mysterious object in GW190814, may indeed consist of deconfined quark matter rather than unusually light black holes. Ultimately, our results suggest that the evolution of magnetized quasiparticle degrees of freedom at high densities is a critical mechanism for reconciling non-perturbative QCD with the latest observations, positioning the thermalized SQS as a compelling candidate for the universe's most massive compact objects.

\section*{Acknowledgements}
The authors would like to thank the Department of Physics, Hansraj College, University of Delhi for providing the necessary facilities and a conducive research environment to carry out this work.

\end{document}